# Dynamical switching of lasing emission by exceptional point modulation in coupled microcavities

Yicong Zhang, [1] Weiwei Liu, [1,*] Qingjie Liu,[1] Bing Wang,[1] and Peixiang Lu[1,2,*]

[1]*Wuhan National Laboratory for Optoelectronics and School of Physics, Huazhong University of Science and Technology, Wuhan 430074, China*
[2]*Hubei Key Laboratory of Optical Information and Pattern Recognition, Wuhan Institute of Technology, Wuhan 430205, China*
*\*Email: lwhust@hust.edu.cn, lupeixiang@hust.edu.cn*

**Abstract:** In a non-Hermitian optical system with loss and gain, an exceptional point (EP) will arise under specific parameters where the eigenvalues and eigenstates exhibit simultaneous coalescence. Here we report a dynamical switching of lasing behavior in a non-Hermitian system composed of coupled microcavities by modulating the EPs. Utilizing the effect of gain, loss and coupling on the eigenstates of coupled microcavities, the evolution path of the eigenvalues related to the laser emission characteristics can be modulated. As a result, the lasing emission property of the coupled cavities exhibits an dynamical switching behavior, which can also be effectively controlled by tuning the gain and loss of the cavities. Moreover, the evolution behavior in a more complicated system composed of three coupled microcavities is investigated, which shows a better tunability compared with the two-microcavity system. Our results have correlated the EPs in non-Hermitian system with lasing emission in complex microcavity systems, which shows great potential for realizing dynamical, ultrafast and multifunctional optoelectronic devices for on-chip integrations.

## 1. Introduction

Optical systems based on nano-scale devices such as waveguides, optical gratings and optical cavities have aroused extensive research for the rich physical properties and functionalities [1-5]. The systems with gain and loss can be described by a non-Hermitian matrix featuring complex eigenvalues and non-orthogonal eigenstates [6-8]. The eigenvalues of such non-Hermitian systems would experience evolution by tuning one or more variable physical parameters. Interestingly, the real and imaginary parts of the eigenvalues will coalesce simultaneously when parameters reach some specific values, giving rise to exceptional points (EPs) in the parametric space[9]. The EPs will result in a series of interesting and counterintuitive optical effects such as chiral mode conversion [10-13], unidirectional light propagation [14, 15], light pulse stopping [16, 17] and sensing enhancement [18], which shows great promise for the applications in information processing, high-sensitive detection and topological photonics.

    Since EPs in non-Hermitian systems were found to be useful for modulating the laser properties, utilizing parametric evolution in the vicinity of EP has attracted great attentions in laser technology, which shows great promise for realizing high-quality and multifunctional lasing devices. For examples, time-asymmetric topological mode transfer has been experimentally realized in silicon photonic device by dynamically encircling an EP [12], and counterintuitive loss-induced enhancement of intracavity field intensity has been observed in coupled microresonators [19]. More recently, controlled laser emission based on pump-induced EPs has been demonstrated in coupled ridge and microdisk cavities [20, 21]. Moreover, lasing modes switching realized by gain and loss modulation is a typical application based on the dynamic parametric evolution of a non-Hermitian system. Although earlier work has suggested the evolution behavior in an optical resonator system with variable pump strength

[20], the effect of other parameters modulation such as coupling strength and more complicated ones still remains unexplored. Therefore, further research is required to clarify the evolution mechanism of multiple parameters to realize more effective lasing emission switching based on complex microcavity systems.

In this work, we theoretically study the qualitative lasing response in a non-Hermitian system consisted of coupled microcavities. The evolution of eigenvalues for the non-Hermitian matrix describing such a system is monitored through continuous tuning of the gain and loss of the microcavities. The inversion of the evolution path occurs at EP marks the reversal of the gain and loss of the corresponding laser behavior, which indicates that the lasing mode can be turned off again as the gain of the system increases. Therefore, dynamical switching of the lasing modes is realized by tuning loss and gain of the coupled microcavities around EP. We also studied the evolution behavior in a more complicated system composed of three coupled microcavities, which shows a better tunability compared with the two-microcavity system. In addition, numerical simulations were performed to verify the theoretical results.

## 2. Results and discussion

First, a coupled microcavity system consisted of two microdisk resonators with variable gain and loss are proposed, as shown in Figure 1(a). Gain, loss and the distance of the two cavities can be tuned individually. EP can be reached when the gain and loss of the coupled cavities satisfy a specific relation. Earlier studies have shown that the resonant levels in optical resonators can be described by a 2×2 non-Hermitian matrix [20]

$$H = \begin{pmatrix} a+i\alpha & \kappa \\ \kappa & b+i\beta \end{pmatrix}, \qquad (1)$$

in which the diagonal elements represent the resonant levels with eigenfrequencies $a$ and $b$ in the absence of coupling [22]. The effective gain and loss of the two resonant levels are respectively expressed as $\alpha$ and $\beta$. The off-diagonal elements $\kappa$ represents the coupling strength. The eigenvalues of this matrix can be solved as

$$\lambda_1 = \frac{(a+b)+i(\alpha+\beta)+\sqrt{[(a-b)+i(\alpha-\beta)]^2+4\kappa^2}}{2}, \qquad (2)$$

$$\lambda_2 = \frac{(a+b)+i(\alpha+\beta)-\sqrt{[(a-b)+i(\alpha-\beta)]^2+4\kappa^2}}{2}. \qquad (3)$$

Assuming that the two levels have identical frequencies for simplicity ($a = b$). In the initial situation without external pump, both of the levels experience strong net loss expressed by equal negative $\alpha$ and $\beta$. As the pump power increases, the two resonant levels experience tunable loss and gain. Figure 1(b) and 1(c) respectively plot the real part and imaginary part of the two eigenvalues as a function of $\alpha$ and $\beta$. The real parts of the eigenvalues represent the emission frequency of the eigenmodes, and the imaginary parts of eigenvalues represent the gain and loss of the two cavities. In particular, in the parameter region at $|\alpha-\beta| = 2\kappa$, not only the real parts and imaginary parts of the eigenvalues but also the two eigenvectors coalesce simultaneously, indicating the rising of the exceptional points [23, 24]. The EPs will result in the identical emission frequency and gain/loss in both cavities, leading to a coalescence of lasing modes.

As the eigenvalues are strongly dependent on $\alpha$ and $\beta$, the performance of the coupled cavity system will show great difference by tuning the loss and gain. The laser mode is switched on when the evolution path of the eigenvalue passes through the isosurface at Im($\lambda$) = 0, and the modes can be excited only when the imaginary part of the eigenvalue is positive. To make it clear, a contour line (red curve) is added to Figure 1(c) to present the points where Im($\lambda_{1,2}$) equals to zero, and Figure 1(d) shows the projection of the contour line on the $\alpha$-$\beta$ plane. One can observe that the parametric plane is divided into four areas, which represents distinct emission states of the two microcavities. For example, in area I, both cavities can be excited for lasing because the imaginary parts of the eigenvalues are larger than 0. In contrast, both cavities can not be excited due to negative imaginary parts of the eigenvalues in area III. Interestingly, only one cavity can be excited in area II and area IV, indicating that the laser is switchable by utilizing a tunable external pump to change the loss and gain of the two cavities individually.

To make a clear manifestation of the evolution path of the eigenvalues, the external pump is only applied on cavity B to tune the gain, while cavity A keeps a net loss ($\alpha = -0.4$). Figure 2(a)-2(c) show the real part and imaginary part of the eigenvalues ($\lambda_1$ and $\lambda_2$) as a function of $\beta$, with the coupling strength $\kappa = 0.2$, 0.4 and 0.5 respectively. Figure 2 also shows that as the coupling strength $\kappa$ increases, the imaginary parts of the eigenvalues at EP will increase, while the real parts keep constant. Specially, for a strong coupling at $\kappa = 0.5$ (Figure 2(c)), the evolution is composed of three typical processes: (1) Both cavities can not be excited for lasing at first, due to the intrinsic loss of the two cavities. (2) As the value of $\beta$ is continuously increased, the eigenvalues approach to each other gradually and coalesce at EP. As the imaginary parts of the eigenvalues are larger than 0, both of the two cavities can be excited for lasing at EP although cavity A has a net loss. (3) After passing EP, the eigenvalue evolutions of the two cavities experience abrupt change, implying distinct lasing property of the two cavities. Interestingly, as $\beta$ further increases, cavity A will be turned off and cavity B will be kept for lasing, which can be explained as the strong mode mismatch between the two cavities and asymmetric light distribution as the gain/loss contrast $|\beta-\alpha|$ is increased beyond $2\kappa$ [25], thus resulting in a decay of the energy in cavity A below the lasing threshold. Figure 2(d) summarizes the dependence of Im($\lambda_{EP}$) on the coupling strength $\kappa$. As $\kappa$ increases over 0.4, the lasing emission in cavity A will experience a dynamical switching process, which corresponds to the situation (3) analyzed above.

Numerical simulation is also carried out to demonstrate the dynamical switching of the lasing emission in coupled cavities. Considering the practical application in photonic devices [26-32], the operation wavelength is selected to be 1550 nm, which corresponds to the communication band. The diameter of the microcavities is set to be 3 μm. Similar to the theoretical calculation above, cavity A has a net loss ($\alpha = -0.4$), and the value of $\beta$ variable to simulate loss and gain. Figure 3(a)-3(c) plot the simulated eigenmodes of the coupled-cavity system with a distance of 300nm, for which the imaginary part of the eigenvalues at EP is below 0 Figure 3(a) shows that the resonant mode has a strong loss when the two cavities have intrinsic loss. As a result, the optical energy will decay rapidly and both the cavities can not be excited for lasing. As the gain of cavity B increases, cavity B can be switched on for lasing while cavity A is never excited, as shown in Figure 3(b) and 3(c). We also increase the cavity coupling by decreasing the distance of the two cavities to 100 nm. In this case, the resonant mode still can not be excited when the two cavities have an intrinsic loss [Figure 3(d)]. However, near the EP, both of the two cavities can be switched on for stable lasing emission [Figure 3(e)]. Moreover, as the gain of cavity B further increases, cavity A will be turned off again [Figure 3(f)]. Therefore, we can conclude that the theoretical analysis is in good agreement with the numerical simulation, which indicates that a dynamical switching of the lasing emission in the coupled-cavity system is demonstrated based on EPs.

The physical mechanism can also be extended to more complicated systems such as three coupled cavities, as schemed in Figure 4(a). Similarly, the non-Hermitian matrix for the coupled-cavity system can be expressed as

$$H = \begin{pmatrix} a+i\alpha & \kappa_1 & 0 \\ \kappa_1 & b+i\beta & \kappa_2 \\ 0 & \kappa_2 & c+i\gamma \end{pmatrix}, \quad (4)$$

which is able to support a third-order EP with appropriate parameter settings [33-35]. Considering the evolution paths of the eigenvalues are more complicated than those in the systems consisted of two cavities, the eigenfrequencies of the three cavities are assumed to be equal ($a = b = c = 1$) for simplification. The cavities A and C experience balanced gain and loss ($\alpha = -\gamma = 0.4$), and the loss/gain of cavity B is tunable. Firstly we consider the situation of identical coupling strength between adjacent cavities ($\kappa_1 = \kappa_2 = \kappa$). In the case of weak coupling, two second-order intersection points with equivalent real parts exist. As the coupling strength increases, these two intersection points get close to each other gradually and merge into a third-order EP at $\beta = 0$ when $\kappa$ reaches 0.2828. As $\kappa$ continues increasing, the EP disappears instantly and the three paths will no longer intersect. The evolutions of the calculated eigenvalues ($\lambda_1$, $\lambda_2$, $\lambda_3$) for the three-cavity system as a function of $\beta$ are plotted in Figure 4(b)-4(d), for $\kappa=0.25$, 0.2828 and 0.3 respectively. Cavity A can be continuously kept for lasing as the eigenvalue corresponding to A always possesses a positive imaginary part. As $\beta$ increases over 0, cavity B will be excited for lasing as well because the imaginary part of the eigenvalue corresponding to B is also above 0, resulting in effective gain in cavity B.

However, the evolution process is different under the circumstance with unequal $\kappa_1$ and $\kappa_2$. Figure 4(e)-4(g) show the evolution paths of the eigenvalues and their projection with $\kappa_2 = 0.36$, 0.375 and 0.4 respectively ($\kappa_1 = 0.2$). One can observe that the evolution paths of the eigenvalues experience distinct processes as $\kappa_2$ increases, with two intersection points [Figure 4(e)], one third-order EP [Figure 4(f)] and no intersection point [Figure 4(g)] respectively. Compared with the previous circumstance, the eigenvalue at the third-order EP possesses a positive imaginary part, leading to a process that all the three eigenvalues possess positive imaginary parts simultaneously. This implies that all of the cavities can be excited simultaneously in the case of $\kappa_1 \neq \kappa_2$. For a specific value $\kappa_2 = 0.375$, the evolution is composed of three processes. (1) Only cavity A can be excited for lasing initially. (2) As $\beta$ increases and three eigenvalues get closer gradually near the third-order EP, the imaginary parts of all three eigenvalues are above 0 and all cavities can be excited for lasing even though cavity C remains net-loss. (3) Three eigenvalues separate after passing EP and as $\beta$ is continuously increased, cavity C is turned off while cavity A and B keep for lasing. The reversal of emission state in cavity C indicates the realization of dynamical switching of the lasing emission based on third-order EP in the system consisted of three cavities.

We have also performed analogous simulation using the same wavelength and microcavity parameters to demonstrate the dynamical switching process. The distance between adjacent cavities is set to be $d_1 = d_2 = d = 50$ nm to simulate an equivalent coupling strength between the adjacent microcavities. The simulated results are plotted in Figure 5(a)-5(c). Figure 5(a) shows that only cavity A can be excited for lasing, because there is a net gain in cavity A, while an identical loss in cavity B and C. As $\beta$ increases to be positive cavity B can be switched on, and cavity C still remains off, as shown in Figure 5(b) and 5(c). We also investigate the influence of the cavity distance on the evolution process of the cavity emission, in which the distances are set to be $d_1 = 95$ nm and $d_2 = 50$ nm to produce an unequal coupling strength between the adjacent microcavities, as shown in Figure 5(d)-5(f). Different from the situation of $d_1 = d_2$, all of the microcavities can be switched on in the vicinity of EP with $\beta > 0$ in this case, as shown in Figure 5(e). As the gain of cavity B further increases, cavity C will be turned off again [Figure 5(f)]. The simulation results fit our theoretical analysis well, indicating a demonstration of the third-order-EP-based dynamical switching of the lasing emission. Compared with the two-microcavity system, the three-microcavity system can support richer evolution process, which shows a better tunability of the emission states in the coupled-microcavity system.

## 3. Conclusion

In conclusion, a dynamical switching of the lasing modes were realized by modulating the exceptional point in a non-Hermitian system composed of coupled microcavities. The evolution of the eigenvalues for the non-Hermitian system was systematically investigated by tuning the gain, loss and coupling strength of the coupled cavities, which provides a convenient approach for dynamical controlling the lasing emission in the cavities. Specifically, we studied the evolution behavior in a more complicated system composed of three coupled microcavities, which shows a better tunability of the emission states compared with the traditional two-microcavity system. Moreover, numerical simulations were also performed and the results can well verify the theoretical analysis. The dynamical modulation of lasing modes utilizing EPs in non-Hermitian system would pave the way for implementation of versatile functionalities in laser technology, which shows great promise for realizing multifunctional devices for optoelectronic integrations

**Funding**
National Natural Science Foundation of China (11804109, 11674117); Doctoral Fund of Ministry of Education of China (20130142110078).

**Disclosures**
The authors declare no conflicts of interest.

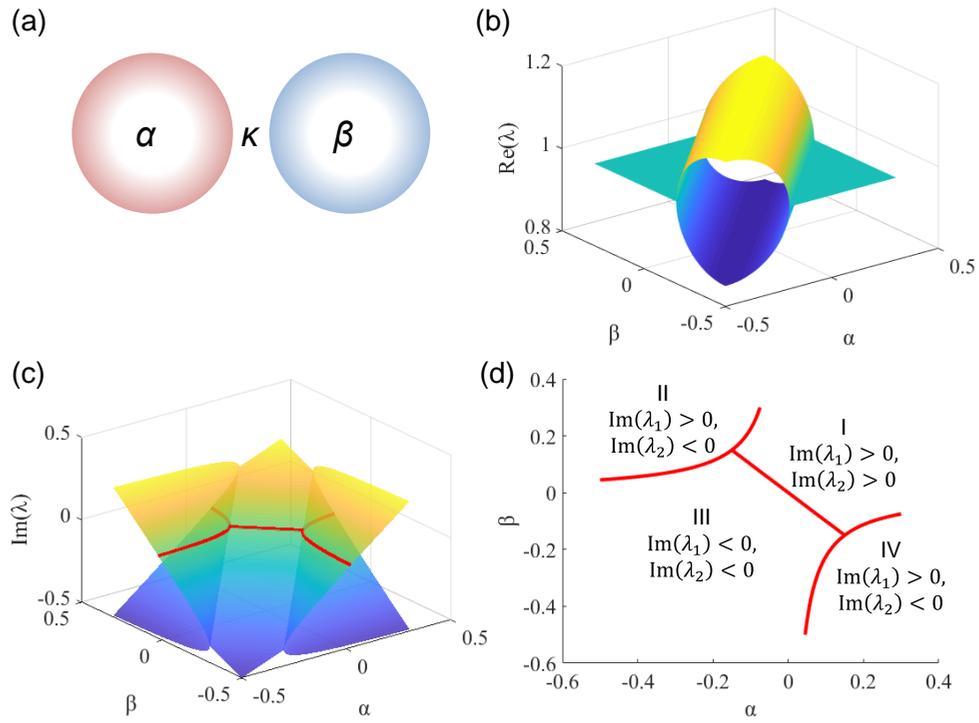

Fig. 1. (a) Schematic of a system composed of two coupled microcavities. The coupling strength is $\kappa$, and the gain/loss is expressed as $\alpha$ and $\beta$ for each cavity respectively. Plots of (b) real part and (c) imaginary part of the eigenvalues as a function of $\alpha$ and $\beta$ ($a = b = 1$, $\kappa = 0.15$). The red curve represents the lasing threshold where Im($\lambda$) = 0. (d) The projection of the zero contour line in (c) on the $\alpha$-$\beta$ parameter plane.

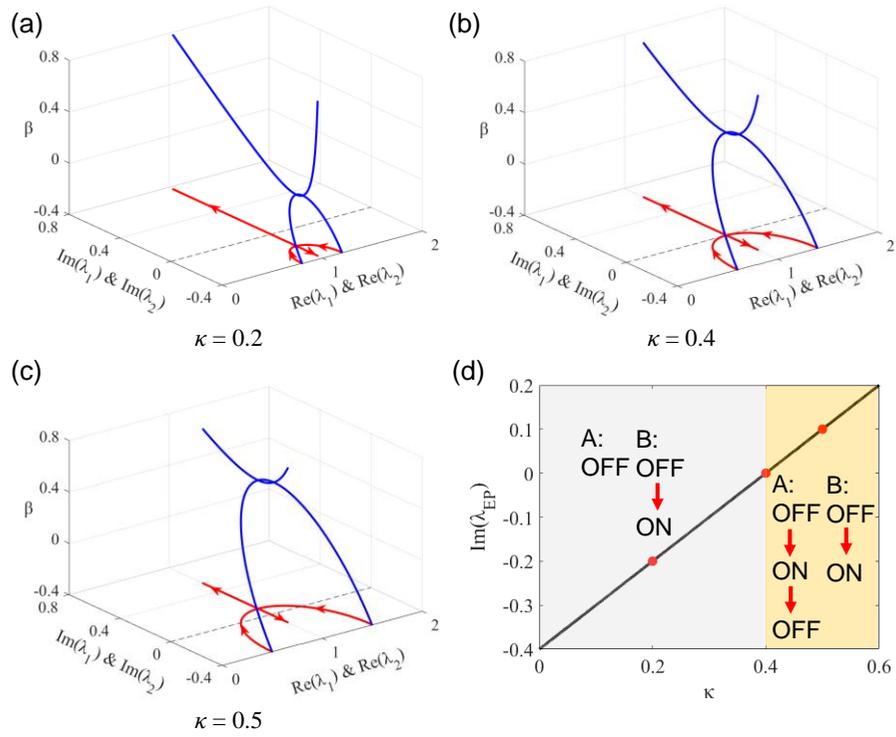

Fig. 2. Plots of the real part and imaginary part of the eigenvalues ($\lambda_1$ and $\lambda_2$) as a function of $\beta$, for (a) $\kappa = 0.2$, (b) $\kappa = 0.4$ and (c) $\kappa = 0.5$ respectively. The red curves represent the projection of the corresponding blue curve on the complex plane of eigenvalues. The black dashed line labels the lasing threshold. (d) Plot of Im($\lambda_{EP}$) as a function of the coupling strength $\kappa$.

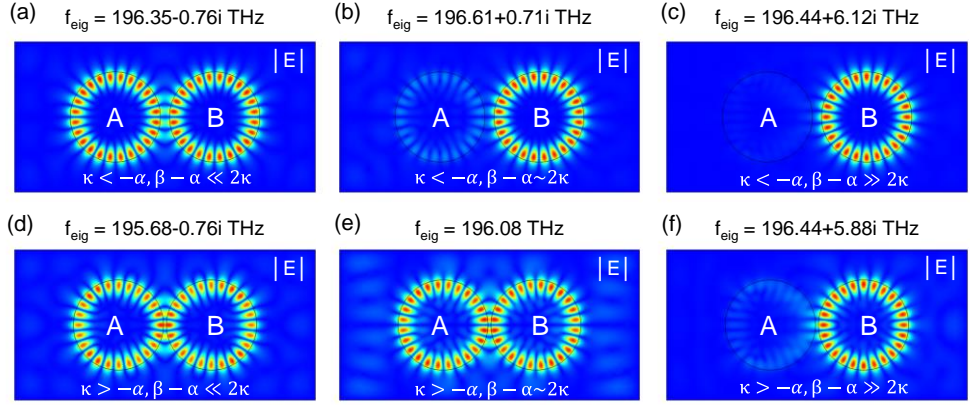

Fig. 3. Simulated eigenmodes and the corresponding eigenfrequencies of the non-Hermitian system consisted of two coupled microcavities for various gain of cavity B. (a)(d) $\beta \ll -\alpha$, (b)(e) $\beta = -\alpha$, (c)(f) $\beta \gg -\alpha$. The adjacent distance between the cavities are set as (a)-(c) $d = 300$ nm ($\kappa < -\alpha$), and (d)-(f) $d = 100$ nm ($\kappa > -\alpha$).

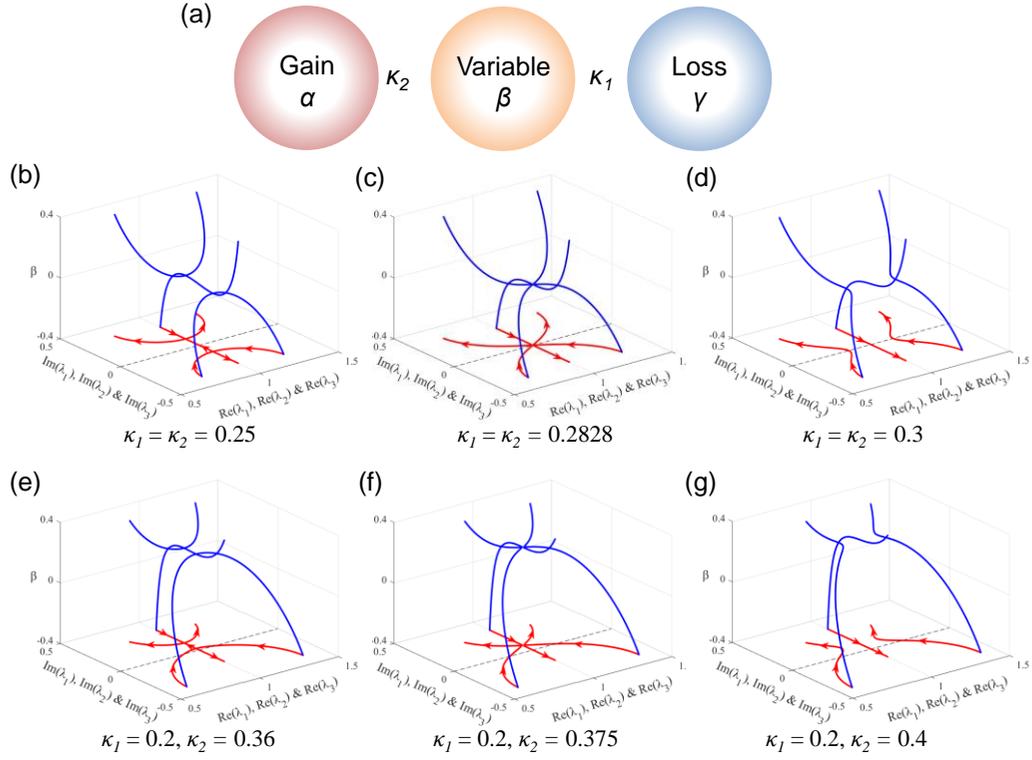

Fig. 4. (a) Schematic of a system composed of three coupled microcavities. The gain/loss of cavities A and C are set to be $\alpha = 0.4$ and $\gamma = -0.4$, respectively. The gain/loss of cavity B can be varied from -0.4 to 0.4. The coupling strengths between adjacent cavities are represented by $\kappa_1$ and $\kappa_2$ respectively. Plots of the real part and imaginary part of the eigenvalues ($\lambda_1$, $\lambda_2$ and $\lambda_3$) as a function of $\beta$ for an equal coupling strength between the adjacent cavities at (b) $\kappa_1 = \kappa_2 = 0.25$, (c) $\kappa_1 = \kappa_2 = 0.2828$ and (d) $\kappa_1 = \kappa_2 = 0.3$ respectively. Plots of the real part and imaginary part of the eigenvalues ($\lambda_1$, $\lambda_2$ and $\lambda_3$) as a function of $\beta$ for an unequal coupling strength between the adjacent cavities at (e) $\kappa_1 = 0.2$, $\kappa_2 = 0.36$, (f) $\kappa_1 = 0.2$, $\kappa_2 = 0.375$ and (g) $\kappa_1 = 0.2$, $\kappa_2 = 0.4$ respectively.

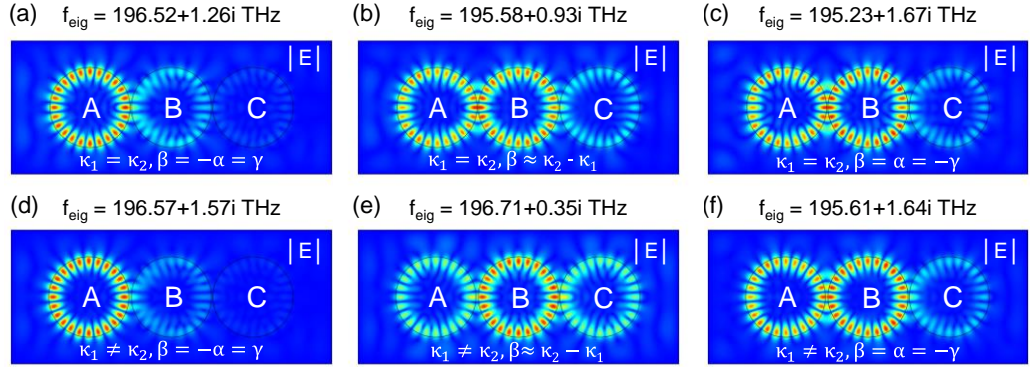

Fig. 5. Simulated eigenmodes and the corresponding eigenfrequencies of the non-Hermitian system consisted of three coupled microcavities. Cavities A and C have balanced gain and loss ($\alpha = -\gamma$), and the gain/loss of cavity B is set as (a)(d) $\beta = -\alpha = \gamma$, (b)(e) $\beta \approx \kappa_2 - \kappa_1$, and (c)(f) $\beta = \alpha = -\gamma$. The distance between adjacent cavities is set to be (a)-(c) $d_1 = d_2 = d = 50$ nm ($\kappa_1 = \kappa_2$), and (d)-(f) $d_1 = 95$ nm, $d_2 = 50$ nm ($\kappa_1 \neq \kappa_2$), respectively.